\documentclass[10pt, conference, letterpaper]{IEEEtran}

\IEEEoverridecommandlockouts

\usepackage{cite}
\usepackage{amsmath,amssymb,amsfonts}
\usepackage{mathptmx} \usepackage{amsthm}
\usepackage{algorithm}
\usepackage{tcolorbox}
\usepackage{url}
\usepackage{graphics} \usepackage{epsfig} \usepackage{enumerate}
\usepackage{booktabs}
\usepackage[bottom]{footmisc}
\usepackage[font=footnotesize,skip=1pt]{caption}

\usepackage{enumitem}
\setlist[enumerate]{topsep=0pt,itemsep=-1ex,partopsep=1ex,parsep=1ex,leftmargin=2ex}
\setlist[itemize]{topsep=0pt,itemsep=-1ex,partopsep=1ex,parsep=1ex,leftmargin=2ex}

\usepackage{lettrine}
\usepackage{verbatim}

\usepackage{graphicx}   \usepackage{subcaption} 
\usepackage{algorithmic}
\usepackage{graphicx}
\usepackage{textcomp}
\usepackage{xcolor}
\def\BibTeX{{\rm B\kern-.05em{\sc i\kern-.025em b}\kern-.08em
    T\kern-.1667em\lower.7ex\hbox{E}\kern-.125emX}}

\newtheorem{theorem}{Theorem}
\newtheorem{corollary}[theorem]{Corollary}

\newtheorem{remark}[theorem]{Remark}

\newcommand{\AoI}{\Delta}
\newcommand{\timeStamp}{\tau}

\newcommand{\interDeliv}{I}

\begin{document}

\title{Analysis and Optimization of Age of Information in Large-Scale Random Access Networks}

\author{
\IEEEauthorblockN{
Yueqi Chen\IEEEauthorrefmark{1},
Siqi Fan\IEEEauthorrefmark{2},
Zhuoyi Zhao\IEEEauthorrefmark{3},
I-Hong Hou\IEEEauthorrefmark{2}, and
Igor Kadota\IEEEauthorrefmark{1}\IEEEcompsocitemizethanks{\IEEEcompsocthanksitem This work was supported in part by NSF grant CNS-2433807 and NSF grant CCF-2332800.}}
\IEEEauthorblockA{
\IEEEauthorrefmark{1}\textit{Northwestern University}
\quad
\IEEEauthorrefmark{2}\textit{Texas A\&M University}
\quad
\IEEEauthorrefmark{3}\textit{University of Hong Kong}}
}

\maketitle
\begin{abstract}

We consider a large-scale network where multiple nodes transmit time-sensitive information via a shared channel to a wireless base station using a Carrier-Sense Multiple Access (CSMA) protocol. In this network, packet transmissions from different nodes are interdependent, making it challenging to analyze and optimize information freshness, captured by the Age of Information (AoI) metric.

In this paper, we develop a tractable analytical framework for analyzing and optimizing the AoI. By combining mean-field approximation with first return time and renewal reward analysis, the proposed approach provides a fast, tractable, and accurate AoI approximation that explicitly accounts for carrier sensing, enabling us to quickly find the network parameters that minimize AoI. Since the framework is built on a finite-state Markov representation of the CSMA protocol, it applies to a broad class of Random Access systems. As a concrete example, we study age-threshold CSMA protocols and derive closed-form or efficiently computable AoI approximations under different protocol settings. We validate the proposed AoI analysis via extensive simulations, demonstrating a tight agreement between analysis and simulation.
\end{abstract}

\section{Introduction}\label{sec6.Intro}
The Age of Information (AoI) quantifies the freshness of information at a receiver by measuring the time elapsed since the generation of the most recently received update~\cite{6195689,kaul2011minimizing}. It has emerged as a key performance metric for wireless systems in which timely information is essential. In many practical systems, information updates are delivered over Random Access networks, particularly those employing CSMA protocol, owing to the low cost, technological maturity, and widespread availability of CSMA-based technologies such as Wi-Fi. Representative applications include warehouse monitoring robots deployed by Amazon~\cite{Kiva,KivaWiFi} and Alibaba~\cite{Alibaba}, vehicle collision-avoidance systems~\cite{DSRC,PerformanceDSRC,SimVsRealDSRC,VehicleToInfrastructureDSRC,AppLevelDSRC}, and path planning and motion control for formations of aerial robots~\cite{Drones,DistributedRobotFormation} and ground robots~\cite{RobotFormation}. These practical deployments motivate the study of AoI performance in CSMA-based Random Access networks.

Despite their practical advantages, Random Access networks face significant performance challenges as the network size and traffic load increase. With more nodes competing for channel access, the network may become congested or overloaded, leading to longer update delays and a sharp degradation in information freshness. Optimizing AoI is therefore particularly important in large-scale Random Access networks. However, the AoI analysis of CSMA protocols is challenging because carrier sensing and acknowledgment (ACK) mechanisms introduce dependencies among nodes. Specifically, the transmission decision of each node depends on recent channel activity and the transmission outcomes of other nodes. These interdependencies couple the state evolution across the network, causing the joint state space to grow rapidly with the number of nodes and making exact joint-state analysis computationally intractable for large-scale networks.

\vspace{0.5ex}
\noindent\textbf{Our contributions.}
The main contributions of this paper are summarized as follows:
\begin{itemize}
    \item We develop a framework for analyzing AoI in large-scale Random Access networks operating under a broad class of CSMA protocols. The framework models a tagged source as a finite-state Markov process coupled with the other sources through a mean field, thereby avoiding the explicit characterization of the joint network state and enabling a tractable approximation of the average AoI.

    \item We apply the proposed framework to age-threshold CSMA as a concrete example. Using first-return-time and renewal-reward analysis, we derive closed-form expressions when available and provide efficient numerical procedures for computing the approximate average AoI otherwise.

    \item We use the proposed analysis to optimize the protocol parameters and validate the analytical results through extensive simulations. The analytical approximations closely match the simulation results across the considered network configurations while requiring substantially less computation than direct simulation.
\end{itemize}

\vspace{0.5ex}
\noindent\textbf{Related work.}
Existing approaches to the analysis and optimization of AoI in Random Access networks can be broadly classified into three categories: (i) protocol-specific analysis, (ii) stochastic hybrid systems (SHS)-based analysis, and (iii) simulation-driven evaluation.

Protocol-specific studies, such as
\cite{10138556,10323421,8493069}, analyze particular Random
Access mechanisms and age-aware protocols by exploiting their
specific structural properties. Related analytical approaches
include the mean-field and second-order framework for
single-slot age-threshold Random Access in
\cite{10943050} and the renewal-based analysis of Random
Access networks with stochastic arrivals in
\cite{9488897}. Such analysis can yield closed-form AoI
expressions, provide insight into network behavior, and
facilitate protocol parameter optimization. However, the
analytical techniques and resulting expressions are generally
tailored to the protocol and assumptions under consideration,
which limits their applicability to other CSMA
configurations.

SHS-based studies, including
\cite{8469047,10180058}, model the joint evolution of discrete
network states and continuous age processes within a unified
stochastic framework. This approach can capture complex
dependencies among queues, transmissions, and AoI. However,
its state space dimension generally grows rapidly with the
network size. Consequently, exact SHS analysis is typically
tractable only for small or highly structured networks, and
explicit AoI expressions can be difficult to obtain.

Simulation-driven studies, such as
\cite{7357587,kaul2011minimizing}, use network simulators
implemented in platforms such as MATLAB and Python, or network
emulators such as CORE~\cite{7357587} and EMANE, to evaluate
average AoI under different network configurations.
Simulation and emulation can accommodate detailed and
analytically intractable protocol behavior, but they may
require substantial computational effort, particularly when
evaluating large networks or searching over a large
protocol-parameter space.

In contrast to these approaches, the proposed mean-field framework provides a tractable approximation of AoI without explicitly modeling the joint state of the entire network. In Sec.~\ref{sec.iv}, we apply the framework to the analysis and optimization of symmetric age-threshold CSMA networks with arbitrary packet lengths and pause durations. In Sec.~\ref{sec6.simulation1}, we compare the computational cost of the proposed approach with that of simulation-based AoI evaluation and optimization.

The remainder of this paper is organized as follows. Sec.~\ref{sec6.model} presents the network model. Sec.~\ref{sec6.analysis} introduces the proposed AoI analysis framework. Sec.~\ref{sec.iv} applies the framework to age-threshold CSMA and develops algorithms for computing the average AoI and optimizing the protocol parameters. Sec.~\ref{sec6.simulation1} presents numerical results that validate the proposed analysis against simulation. Finally, Sec.~\ref{sec6.summary} concludes the paper.

\section{Network Model}\label{sec6.model}
Consider a single-hop Random Access wireless network with \(N\) sources transmitting time-sensitive information to a base station (BS) according to an age-threshold CSMA protocol. We assume that sources can generate fresh packets on demand (commonly known as the generate-at-will model) and, after each successful packet transmission to the BS, ``pause'' for a deterministic period of time before attempting to transmit again (commonly known as the age-threshold CSMA protocol~\cite{10943050}). 

\vspace{0.5ex}
\noindent\textbf{Age-threshold CSMA Protocol.}
Time is slotted into \emph{mini-slots} of duration \(\delta\) seconds,
indexed by \(k=1,2,\ldots\), where \(\delta\) is the time required for sources to detect an ongoing transmission. A \emph{transmission opportunity}
occurs when a source is neither transmitting nor in its pause period
and senses the channel to be idle during a mini-slot. At each transmission opportunity,
source \(i\) attempts transmission with probability \(p\in(0,1)\);
otherwise, it remains idle. Prior to attempting transmission, source \(i\) generates a fresh packet. 
If source \(i\) is the only source attempting transmission during
mini-slot \(k\), its transmission is successful. All other sources
detect the transmission by the beginning of mini-slot \(k+1\) and defer
until it is completed. If two or more sources attempt transmission in
the same mini-slot, a \emph{packet collision} occurs, and no packet is
successfully received. After a collision, the sources discard the
collided packets and generate fresh packets before their next
transmission attempt. Each transmission attempt, successful or not, occupies the channel for \(L\) mini-slots, where \(L\) denotes the packet length measured in mini-slots.

After a successful transmission, the transmitting source enters
a \emph{pause period} of exactly \(H\) mini-slots, during which
it does not attempt a new transmission. It becomes eligible to
contend for the channel again after the pause period. Since its
AoI is reset to \(L\) upon delivery, the pause length \(H\)
corresponds to an AoI contention threshold of \(L+H\)
mini-slots. We assume that there are no hidden or exposed
sources and that feedback from the BS is instantaneous and
error-free.

\vspace{0.5ex}
\noindent\textbf{Markov Chain Model.}
The joint evolution of the sources' states forms a finite-state Markov chain.
To avoid explicitly modeling the joint state, we represent a tagged
source by a reduced state process whose transition probabilities capture
the interactions with other sources through the mean-field approximation
developed in Sec.~\ref{sec6.analysis}. We next define its action
space, state space, and transition dynamics.

\vspace{0.5ex}
\noindent\underline{Action space.}
Sources continually sense the channel and track each others transmissions. At the beginning of mini-slot $k$, if the tagged source is not in its pause period, is not transmitting, and is not expecting transmissions from other sources, then it obtains a transmission opportunity and takes one of two possible actions: (i) \emph{Transmit}, with probability \(p\), or (ii) \emph{Not Transmit}, with probability \(1-p\). The action space at each transmission opportunity is represented by \(\mathcal{A}=\{\text{Transmit},\ \text{Not Transmit}\}.\)
The source does not know whether another source will attempt transmission in the same mini-slot when choosing its action.

\vspace{0.5ex}
\noindent\underline{State space.}
The state of a tagged source \(i\) at the beginning of mini-slot \(k\) describes its current transmission status, the remaining duration of any ongoing channel occupation, and the remaining duration of its pause period, when applicable. The state belongs to one of the following categories:

\begin{itemize}

\item \emph{Idle state:} Occurs when the tagged source $i$ is not in its pause period, is not transmitting, and is not expecting transmissions from other sources at the beginning of mini-slot $k$. The Idle state represents a transmission opportunity for the tagged source and for other non-paused sources. 

\item \emph{Transmission states:}
The state is denoted by \(\text{TX}_j\) when tagged source \(i\) is performing a successful transmission and still needs \(j\) mini-slots, including mini-slot \(k\), to complete the transmission. 
These states progress deterministically as
\[
\text{TX}_j \rightarrow \text{TX}_{j-1},
\qquad j=L-1,\ldots,2.
\]
After completing the transmission in state \(\text{TX}_1\), source \(i\) enters the pause period. If \(H=0\), it instead returns to the Idle state with probability $1$.

\item \emph{Busy states:}
The state is denoted by \(\text{Busy}_b\) when tagged source $i$ is not successfully transmitting a packet and expects the channel to remain occupied for \(b\) mini-slots, including mini-slot \(k\). A Busy state represents either a transmission by one or more other sources or a collision involving source \(i\). 
The Busy states progress deterministically as
\[
\text{Busy}_b \rightarrow \text{Busy}_{b-1},
\qquad b=L-1,\ldots,2.
\]
After the channel occupation ends in state \(\text{Busy}_1\), source \(i\) returns to the Idle state with probability $1$.

\item \emph{Pause states:}
The state is denoted by \(\text{Pause}_h\) when tagged source \(i\) is in its post-success pause period with \(h\) mini-slots remaining, including mini-slot \(k\), and no channel occupation is ongoing at the beginning of mini-slot \(k\). While the channel remains unoccupied, the counter decreases by one in every mini-slot.

\item \emph{Pause-Busy states:}
The state is denoted by \(\text{PauseBusy}_{h,b}\) when source \(i\) has \(h\) mini-slots remaining in its pause period and the channel will remain occupied for \(b\) mini-slots, both including mini-slot \(k\). In each mini-slot, both counters decrease simultaneously. If the channel occupation ends before the pause period, the source transitions to the corresponding Pause state. If the pause period ends while the channel remains occupied, the source transitions to the corresponding Busy state. If both end simultaneously, the source transitions to the Idle state.
\end{itemize}

Accordingly, for \(L\geq2\) and \(H\geq0\), the tagged-source state space is
\begin{align}\label{eq:state_space}
    \mathcal{S}_{\mathrm{R}}(L,H)
={}&
\{\mathrm{Idle}\}
\cup
\{\mathrm{TX}_j:1\leq j\leq L-1\}
\nonumber\\
&\cup
\{\mathrm{Busy}_b:1\leq b\leq L-1\}
\nonumber\\
&\cup
\{\mathrm{Pause}_h:1\leq h\leq H\}
\nonumber\\
&\cup
\bigl\{
\mathrm{PauseBusy}_{h,b}:
(h,b)\in\mathcal{R}_{\mathrm{PB}}(L,H)
\bigr\},
\end{align}
where
\[
\begin{aligned}
\mathcal{R}_{\mathrm{PB}}(L,H)
=
\bigl\{(h,b):\;&
1\leq h\leq H-1,\;
1\leq b\leq L-1,
\\[-1mm]
&
1\leq h+L-1-b\leq H-1
\bigr\}.
\end{aligned}
\]
For \(H=0\), both the Pause and PauseBusy state sets are
empty. For \(H=1\), the PauseBusy state set is empty. When
\(L=1\), the TX, Busy, and PauseBusy countdown states are
omitted.

\vspace{0.5ex}
\noindent\underline{Transition dynamics.}
The evolution of the tagged source is determined by its current state,
its transmission decision when eligible, and the simultaneous
transmission decisions of the other sources. The corresponding
transition matrices are constructed in Sec.~\ref{secIII-A}.

\vspace{0.5ex}
\noindent\textbf{Network Age of Information.}
Let \(\timeStamp_i(k)\) denote the generation time of the freshest
packet successfully received by the BS from source \(i\) by the
beginning of mini-slot \(k\). The AoI of source \(i\) at that time is \(\AoI_i(k):=k-\timeStamp_i(k).\) The AoI increases by one in each mini-slot without a successful
delivery and is reset to \(L\) upon a successful delivery. 
The infinite-horizon expected network AoI (NAoI) is
\begin{equation}\label{eq6.NAoI_def}
\mathrm{NAoI}
:=
\lim_{K\rightarrow\infty}
\frac{1}{KN}
\sum_{k=1}^{K}
\sum_{i=1}^{N}
\mathbb{E}\!\left[\AoI_i(k)\right].
\end{equation}

\section{Age of Information Analysis Framework}\label{sec6.analysis}

In this section, we develop a low-complexity framework for
analyzing the AoI of a broad class of CSMA protocols that admit
finite-state Markov chain representations and mean-field
approximations of inter-source interactions. For concreteness,
we focus on the age-threshold CSMA protocol, which includes
conventional CSMA as the special case \(H=0\).

The proposed framework consists of three main steps: (i) constructing the transition matrix of a tagged source under a mean-field approximation of the interactions among sources; (ii) deriving the first and second moments of the interdelivery interval of the tagged source; and (iii) evaluating the AoI using these moments through a renewal-reward analysis.

\subsection{Construction of Transition Matrices}\label{secIII-A}

The first step of the framework is to construct the transition matrix of the tagged-source Markov chain. For the age-threshold CSMA protocol, the state space and the mean-field approximation of the interactions among sources depend on the relationship between the packet length \(L\) and the pause length \(H\). We consider three cases: (i) \(H=0\); (ii) \(0<H\leq L\); and (iii) \(H\geq L+1\). We first consider \(L\geq2\); the special case \(L=1\), in which a transmission attempt is completed within its initiation mini-slot, is treated separately below.

\vspace{0.5ex}
\noindent\textbf{Case (i): \(H=0\).}
This case corresponds to the traditional CSMA protocol without a pause following a successful transmission. Using the state definition in Sec.~\ref{sec6.model}, the state space of the tagged source is 
\[
\mathcal{S}_0
=
\{\mathrm{Idle}\}
\cup
\{\mathrm{TX}_j:1\leq j\leq L-1\}
\cup
\{\mathrm{Busy}_b:1\leq b\leq L-1\}.\]
When the tagged source is in the Idle state, it chooses to transmit with probability \(p\). Since \(H=0\), all other sources are also eligible to contend. Hence, given that the tagged source $i$ in the Idle state, the probability that none of the other \(N-1\) sources attempts transmission is given by \(Q_0=(1-p)^{N-1}.\)
Consequently, from the Idle state, the tagged source remains in Idle with probability \(Q_0(1-p)\), begins a successful transmission with probability \(Q_0p\), and enters a Busy state with probability \(1-Q_0\).

\vspace{0.5ex}
\noindent\emph{Transition matrix.}
The transition matrix \(\mathbf{M}_0=\mathbf{M}_0(Q_0,p,L)\) is defined over \(\mathcal{S}_0\).
Its nonzero entries are
\begin{align}
M_0(\mathrm{Idle},\mathrm{Idle})
&=Q_0(1-p), \notag\\
M_0(\mathrm{Idle},\mathrm{TX}_{L-1})
&=Q_0p, \notag\\
M_0(\mathrm{Idle},\mathrm{Busy}_{L-1})
&=1-Q_0, \notag\\
M_0(\mathrm{TX}_j,\mathrm{TX}_{j-1})
&=1,
&&j=L-1,\ldots,2, \notag\\
M_0(\mathrm{TX}_1,\mathrm{Idle})
&=1, \notag\\
M_0(\mathrm{Busy}_b,\mathrm{Busy}_{b-1})
&=1,
&&b=L-1,\ldots,2, \notag\\
M_0(\mathrm{Busy}_1,\mathrm{Idle})
&=1.
\label{eq:M0_entries}
\end{align}
All remaining entries of \(\mathbf{M}_0\) are zero.

\vspace{0.5ex}
\noindent\textbf{Case (ii): \(0<H\leq L\).}
We next consider the age-threshold CSMA protocol with a short
pause period following a successful transmission. For
\(0<H\leq L\), the tagged source evolves over
\(\mathcal{S}_1:=\mathcal{S}_{\mathrm{R}}(L,H)\), where
\(\mathcal{S}_{\mathrm{R}}(L,H)\) is defined in \eqref{eq:state_space}. 

Because successful-delivery completions are separated by at least
\(L\) mini-slots and \(H\leq L\), at most one non-tagged source
can be paused in any mini-slot. Let \(N_P\in\{0,1\}\) denote
the number of paused non-tagged sources, and define
\(\bar N_P:=\mathbb{E}[N_P\mid S_i=\mathrm{Idle}]\),
where \(S_i\) is the state of the tagged source. The
conditional no-attempt probability is
\begin{align}
Q_1
&=
\mathbb{E}\!\left[
(1-p)^{N-1-N_P}\mid S_i=\mathrm{Idle}
\right]
\notag\\
&=
(1-\bar N_P)(1-p)^{N-1}
+
\bar N_P(1-p)^{N-2}
\notag\\
&=
Q_0+\bar N_Pp(1-p)^{N-2},
\label{eq:Q1}
\end{align}
where \(Q_0:=(1-p)^{N-1}\). Since
\(0\leq\bar N_P\leq1\), the correction term satisfies
\(0\leq\bar N_Pp(1-p)^{N-2}
\leq p(1-p)^{N-2}\).
Therefore, under the large-network scaling \(p=O(1/N)\),
we have \(Q_1=Q_0+O(1/N)\). We use \(Q_1\approx Q_0\)
to obtain a closed-form approximation for the short-pause
regime.

\vspace{0.5ex}
\noindent\emph{Transition matrix.}
Under the approximation \(Q_1\approx Q_0\), the transition matrix
\(\mathbf{M}_1=\mathbf{M}_1(Q_0,p,L,H)\) is defined over
\(\mathcal{S}_1\). Its nonzero entries are
\begin{align}
M_1(\mathrm{Idle},\mathrm{Idle})
&=Q_0(1-p), \notag\\
M_1(\mathrm{Idle},\mathrm{TX}_{L-1})
&=Q_0p, \notag\\
M_1(\mathrm{Idle},\mathrm{Busy}_{L-1})
&=1-Q_0, \notag\\
M_1(\mathrm{TX}_j,\mathrm{TX}_{j-1})
&=1,\quad j=L-1,\ldots,2, \notag\\
M_1(\mathrm{TX}_1,\mathrm{Pause}_H)
&=1, \notag\\
M_1(\mathrm{Busy}_b,\mathrm{Busy}_{b-1})
&=1,\quad b=L-1,\ldots,2, \notag\\
M_1(\mathrm{Busy}_1,\mathrm{Idle})
&=1, \notag\\
M_1(\mathrm{Pause}_h,\mathrm{Pause}_{h-1})
&=Q_0,\quad h=H,\ldots,2, \notag\\
M_1(\mathrm{Pause}_h,\mathrm{PauseBusy}_{h-1,L-1})
&=1-Q_0,\quad h=H,\ldots,2, \notag\\
M_1(\mathrm{Pause}_1,\mathrm{Idle})
&=Q_0, \notag\\
M_1(\mathrm{Pause}_1,\mathrm{Busy}_{L-1})
&=1-Q_0, \notag\\
M_1(\mathrm{PauseBusy}_{h,b},
\mathrm{PauseBusy}_{h-1,b-1})
&=1,\quad h\geq2,\ b\geq2, \notag\\
M_1(\mathrm{PauseBusy}_{h,1},\mathrm{Pause}_{h-1})
&=1,\quad h\geq2, \notag\\
M_1(\mathrm{PauseBusy}_{1,b},\mathrm{Busy}_{b-1})
&=1,\quad b\geq2, \notag\\
M_1(\mathrm{PauseBusy}_{1,1},\mathrm{Idle})
&=1.
\label{eq:M1_entries}
\end{align}
Here, only transitions between states in \(\mathcal{S}_1\) are
included; all remaining entries are zero.

\vspace{0.5ex}
\noindent\textbf{Case (iii): \(H\geq L+1\).}
When the pause duration exceeds the packet length, multiple
non-tagged sources may be paused simultaneously. The tagged
source evolves over
\(\mathcal{S}_2:=\mathcal{S}_{\mathrm{R}}(L,H)\), defined in~\eqref{eq:state_space}.

Let \(\mu_s^{(2)}\) denote the stationary probability of state
\(s\). When no transmission is in progress at the beginning of
a mini-slot, a representative source is either in the Idle
state or in one of the Pause states. The probability that it
is paused under this condition is
\begin{equation}
\pi_P^{(2)}
:=
\frac{
\sum_{h=1}^{H}
\mu_{\mathrm{Pause}_h}^{(2)}
}{
\mu_{\mathrm{Idle}}^{(2)}
+
\sum_{h=1}^{H}
\mu_{\mathrm{Pause}_h}^{(2)}
}.
\label{eq:pause_probability_case2}
\end{equation}

Unlike the short-pause regime, in which the number of paused
non-tagged sources is restricted to zero or one, the
long-pause regime permits multiple non-tagged sources to
remain paused when the channel becomes idle. We therefore
adopt a source-independence mean-field closure. Conditioned on the tagged source being on an Idle state, 
a non-tagged source does not attempt
transmission with probability
\(\pi_P^{(2)}+(1-\pi_P^{(2)})(1-p)
=1-p(1-\pi_P^{(2)})\). Consequently, the probability that none
of the \(N-1\) non-tagged sources attempts transmission is
approximated by
\begin{equation}
Q_2
\approx
\left[
1-p\left(1-\pi_P^{(2)}\right)
\right]^{N-1}.
\label{eq:Q2}
\end{equation}

The transition matrix
\(\mathbf{M}_2=\mathbf{M}_2(Q_2,p,L,H)\) follows the same local
transition rules as \(\mathbf{M}_1\), with \(Q_0\) replaced by
\(Q_2\). Since \(H\geq L+1\), its reachable state space is
generally larger than that of the short-pause case.

The stationary vector
\(\boldsymbol{\mu}^{(2)}
=[\mu_s^{(2)}]_{s\in\mathcal{S}_2}\)
and \(Q_2\) are characterized jointly by
\begin{equation}
\boldsymbol{\mu}^{(2)}
=
\boldsymbol{\mu}^{(2)}
\mathbf{M}_2(Q_2,p,L,H),
\qquad
\boldsymbol{\mu}^{(2)}\mathbf{1}=1,
\label{eq:fixed_point_case2}
\end{equation}
together with \eqref{eq:pause_probability_case2} and
\eqref{eq:Q2}.

\vspace{0.5ex}
\noindent\emph{Special case \(L=1\).}
When \(L=1\), each transmission attempt is completed within
its initiation mini-slot. Hence, the TX, Busy, and PauseBusy
countdown states are omitted. For \(H\geq1\), the tagged-source
state space is \(\mathcal{S}_{L=1}
=
\{\mathrm{Idle}\}
\cup
\{\mathrm{Pause}_h:1\leq h\leq H\}.\)
Let \(Q\) denote the probability that none of the non-tagged
eligible sources attempts transmission. The nonzero transition
probabilities are
\begin{align}
M(\mathrm{Idle},\mathrm{Pause}_H)
&=Qp, \notag\\
M(\mathrm{Idle},\mathrm{Idle})
&=1-Qp, \notag\\
M(\mathrm{Pause}_h,\mathrm{Pause}_{h-1})
&=1,\quad h=2,\ldots,H, \notag\\
M(\mathrm{Pause}_1,\mathrm{Idle})
&=1.
\label{eq:L1_transitions}
\end{align}
The transition
\(\mathrm{Idle}\rightarrow\mathrm{Pause}_H\) marks a
successful packet delivery. When \(H=0\), the only state is
\(\mathrm{Idle}\), and a successful delivery occurs during its
self-transition with probability \(Q_0p\).

\subsection{Interdelivery Interval}\label{secIII.B}
We next analyze the interdelivery interval of a tagged source under the Markov chain constructed in Sec.~\ref{secIII-A}. Let \(\interDeliv_i\) denote the number of mini-slots between two consecutive successful packet deliveries from source \(i\).

For \(L\geq2\), every visit to state \(\mathrm{TX}_1\) corresponds to a successful delivery at the end of the current mini-slot. Therefore, \(\interDeliv_i\) can be represented as the first-return time to the delivery-completion state \(s^\star=\mathrm{TX}_1.\) Because the chain returns to the same state after each successful delivery, the strong Markov property implies that successive first-return times to \(s^\star\) are independent and identically distributed. Hence, the packet delivery epochs form a renewal process under the tagged-source mean-field model.

We denote the first and second moments of the interdelivery interval by \(\mathbb{E}[\interDeliv_i]
\text{ and }
\mathbb{E}[\interDeliv_i^2],\)
respectively. These moments are computed from the corresponding transition matrix \(\mathbf{M}_0\), \(\mathbf{M}_1\), or \(\mathbf{M}_2\) using standard first-return analysis.

\vspace{0.5ex}
\noindent\textbf{Return-time computation.}
Let \(\mathbf{M}\) denote the transition matrix over state space
\(\mathcal{S}\), and let \(s^\star=\mathrm{TX}_1\). For each
\(j\in\mathcal{S}\setminus\{s^\star\}\), define \(u_j
=
\mathbb{E}_j[T_{s^\star}],
v_j
=
\mathbb{E}_j[T_{s^\star}^2],\)
where \(T_{s^\star}\) denotes the first hitting time of \(s^\star\). Set \(u_{s^\star}=
v_{s^\star}=0.\)
The first and second hitting time moments satisfy
\begin{align}
u_j
&=
1+
\textstyle\sum_{k\in\mathcal{S}}
M_{jk}u_k,
\qquad
j\neq s^\star,
\label{eq:return_m1}\\
v_j
&=
1
+
2\textstyle\sum_{k\in\mathcal{S}}
M_{jk}u_k
+
\sum_{k\in\mathcal{S}}
M_{jk}v_k,
\qquad
j\neq s^\star.
\label{eq:return_m2}
\end{align}
Conditioning on the first transition out of \(s^\star\) gives
\begin{align}
\mathbb{E}[\interDeliv_i]
&=
1+
\textstyle\sum_{k\in\mathcal{S}}
M_{s^\star k}u_k,
\label{eq:Ideliv_from_TX1}\\
\mathbb{E}[\interDeliv_i^2]
&=
1
+
2\textstyle\sum_{k\in\mathcal{S}}
M_{s^\star k}u_k
+
\sum_{k\in\mathcal{S}}
M_{s^\star k}v_k.
\label{eq:Ideliv2_from_TX1}
\end{align}

For \(L=1\), no TX countdown state exists. When \(H\geq1\), every
successful delivery is followed by a transition from
\(\mathrm{Idle}\) to \(\mathrm{Pause}_H\). Since
\(\mathrm{Pause}_H\) can be entered only after a successful delivery,
we set \(s^\star=\mathrm{Pause}_H\). The first-return equations
\eqref{eq:return_m1}--\eqref{eq:Ideliv2_from_TX1} then apply with
\(\mathrm{Pause}_H\) as the delivery-marking state. When \(L=1\) and \(H=0\), the only state is \(\mathrm{Idle}\), and a
successful delivery occurs during an
\(\mathrm{Idle}\rightarrow\mathrm{Idle}\) self-transition with
probability \(Q_0p\). In this case, the interdelivery interval is
geometrically distributed and is treated separately in
Sec.~\ref{age-threshold-H=0}.

We evaluate the AoI from the first two interdelivery moments using
renewal-reward analysis. 
\begin{remark}\label{rem:NAoI_general}
For a discrete-time renewal delivery process whose AoI resets
to \(L\) after each successful delivery, specializing~\cite[Theorem~3]{9488897} gives
\begin{equation}
\mathrm{NAoI}
=
L
+
\frac{\mathbb{E}[\interDeliv_i^2]}
{2\mathbb{E}[\interDeliv_i]}
-\frac{1}{2}.
\label{eq6.NAoI}
\end{equation}
\end{remark}

Thus, once the mean-field transition matrix is determined, the NAoI is
obtained by computing the first two interdelivery moments through
\eqref{eq:return_m1}--\eqref{eq:Ideliv2_from_TX1} and substituting them
into \eqref{eq6.NAoI}. In the next section, we apply this procedure to
the three age-threshold CSMA regimes identified in
Sec.~\ref{secIII-A}.

\section{Application of the Framework to Age-threshold CSMA Networks}
\label{sec.iv}
In this section, we apply the framework developed in
Sec.~\ref{sec6.analysis} to the age-threshold CSMA protocol and
develop numerical procedures for computing the stationary distribution,
the interdelivery moments, and the resulting NAoI.

\subsection{Age-threshold CSMA Network with \(H=0\)}
\label{age-threshold-H=0}
When \(H=0\), the age-threshold CSMA protocol reduces to the traditional
CSMA protocol without a post-success pause. The conditional no-attempt probability
is \(Q_0=(1-p)^{N-1},\) and the tagged-source transition matrix is \(\mathbf{M}_0\), as
constructed in Sec.~\ref{secIII-A}.
For \(L\geq2\), the stationary distribution
\(\boldsymbol{\mu}^{(0)}\) is obtained by solving \(\boldsymbol{\mu}^{(0)}
=
\boldsymbol{\mu}^{(0)}\mathbf{M}_0,
\boldsymbol{\mu}^{(0)}\mathbf{1}=1.\)
The stationary probability of the delivery-completion state is
\begin{equation}
\mu_{\mathrm{TX}_1}^{(0)}
=
\frac{p(1-p)^{N-1}}
{L\bigl(1-(1-p)^N\bigr)+(1-p)^N}.
\label{eq:mu_tx1}
\end{equation}

\begin{theorem}
\label{thm:NAoI_H0}
Consider a symmetric CSMA network with packet length \(L\geq1\) and no
post-success pause. The first and second
moments of the interdelivery interval are
\begin{equation}
\mathbb{E}[\interDeliv_i]
=
\frac{
1+(L-1)(1-Q_0+Q_0p)
}{Q_0p},
\label{eq:first_moment_H0}
\end{equation}
and
\begin{equation}
\begin{aligned}
\mathbb{E}[\interDeliv_i^2]
=
\frac{1}{Q_0p}
\Big[
&2\mathbb{E}[\interDeliv_i]-1
+(1-Q_0+Q_0p)(L-1)^2\\
&+2(1-Q_0)(L-1)\mathbb{E}[\interDeliv_i]
\Big],
\end{aligned}
\label{eq:second_moment_H0}
\end{equation}
respectively. Consequently, the exact network AoI is
\begin{equation}
\begin{aligned}
&\mathrm{NAoI}
={}L-\frac{1}{2}
+\frac{1}{2[1+(L-1)(1-Q_0+Q_0p)]}\\
&\times
\Bigg\{
\frac{2[1+(L-1)(1-Q_0+Q_0p)]}{Q_0p}
[1+(1-Q_0)(L-1)]\\
&\hspace{8mm}
-1+(1-Q_0+Q_0p)(L-1)^2
\Bigg\}.
\end{aligned}
\label{eq:NAoI_H0}
\end{equation}
\end{theorem}

\begin{proof}
For \(L\geq2\), solving the first-return equations
\eqref{eq:return_m1}--\eqref{eq:Ideliv2_from_TX1}
for \(\mathbf{M}_0\) yields
\eqref{eq:first_moment_H0} and
\eqref{eq:second_moment_H0}. For \(L=1\), a successful delivery
occurs in each mini-slot with probability \(Q_0p\). Therefore,
\(\interDeliv_i\) is geometrically distributed over
\(\{1,2,\ldots\}\), with \(\mathbb{E}[\interDeliv_i]
=
{1}/{Q_0p} \text{ and }
\mathbb{E}[\interDeliv_i^2]
=
{(2-Q_0p)}/{(Q_0p)^2}.\)
These expressions agree with
\eqref{eq:first_moment_H0} and
\eqref{eq:second_moment_H0} evaluated at \(L=1\).
Substitution into \eqref{eq6.NAoI} yields
\eqref{eq:NAoI_H0}.
\end{proof}

\subsection{Age-threshold CSMA Network with \(0<H\leq L\)}
We next consider the age-threshold CSMA network with a short
pause period. As established in Sec.~\ref{secIII-A}, under
the large-network scaling \(p=O(1/N)\), we use
\(Q_1\approx Q_0=(1-p)^{N-1}\). For \(L\geq2\), the tagged
source evolves over \(\mathcal{S}_1\) with transition matrix
\(\mathbf{M}_1(Q_0,p,L,H)\). The case \(L=1\), for which
\(H=1\), follows from the reduced Markov chain in
\eqref{eq:L1_transitions}. Under this approximation, the
interdelivery moments and NAoI admit the following scalar
expressions.

\begin{theorem}
\label{thm:NAoI_short_pause}
Consider a symmetric age-threshold CSMA network with packet length
\(L\geq1\) and pause length \(0<H\leq L\). Under the large-network
approximation \(Q_1\approx Q_0\), define
\begin{align}
\mathcal{U}
&=
\frac{1+(L-1)(1-Q_0+Q_0p)}{Q_0p},
\label{eq:U_short_pause}\\
\mathcal{V}
&=
\frac{1}{Q_0p}
\Big[
2\mathcal{U}-1
+(1-Q_0+Q_0p)(L-1)^2
\notag\\[-1mm]
&\hspace{16mm}
+2(1-Q_0)(L-1)\mathcal{U}
\Big].
\label{eq:V_short_pause}
\end{align}
Then the first two moments of the interdelivery interval are
\begin{align}
\mathbb{E}[\interDeliv_i]
={}&
\mathcal{U}
+HQ_0^H
+(1-Q_0)
\sum_{t=1}^{H}Q_0^{t-1}(L+t-1),
\label{eq:first_moment_short_pause}\\
\mathbb{E}[\interDeliv_i^2]
={}&
\mathcal{V}
+H^2Q_0^H
+(1-Q_0)
\sum_{t=1}^{H}Q_0^{t-1}(L+t-1)^2
\notag\\
&+
2\mathcal{U}
\left[
HQ_0^H
+(1-Q_0)
\sum_{t=1}^{H}Q_0^{t-1}(L+t-1)
\right].
\label{eq:second_moment_short_pause}
\end{align}
Consequently, the corresponding large-network approximation of NAoI is given by \begin{equation}
{
\widehat{\mathrm{NAoI}}_1
=
L
+
\frac{\mathbb{E}[\interDeliv_i^2]}
{2\mathbb{E}[\interDeliv_i]}
-\frac{1}{2}\;.
}
\label{eq:NAoI_H_leq_L}
\end{equation}
\end{theorem}
\begin{proof}
Applying the first- and second-order return-time equations
\eqref{eq:return_m1}--\eqref{eq:Ideliv2_from_TX1} to
\(\mathbf{M}_1(Q_0,p,L,H)\) and simplifying yields
\eqref{eq:first_moment_short_pause} and
\eqref{eq:second_moment_short_pause}. Substitution into
\eqref{eq6.NAoI} then gives
\eqref{eq:NAoI_H_leq_L}.
\end{proof}

\subsection{Age-threshold CSMA Network with \(H\geq L+1\)}

\noindent\emph{Special case \(L=1\).}
When \(L=1\), the condition \(H\geq L+1\) reduces to
\(H\geq2\). Since each transmission attempt is completed within
one mini-slot, the tagged source alternates between an Idle
period and a deterministic pause of \(H\) mini-slots. For a
fixed conditional no-attempt probability \(Q_2\), a successful
transmission occurs in each Idle mini-slot with probability
\(Q_2p\). Hence, the stationary probability that the tagged
source is paused is
\(\pi_P^{(2)}=HQ_2p/(1+HQ_2p)\). Applying the
source-independence mean-field closure in \eqref{eq:Q2}, the conditional no-attempt probability satisfies
\begin{equation}
Q_2
\approx
\left(
1-\frac{p}{1+HQ_2p}
\right)^{N-1}.
\label{eq:L1_long_pause_fixed_point}
\end{equation}
Thus, \(Q_2\) is obtained by numerically solving
\eqref{eq:L1_long_pause_fixed_point}.

For a fixed solution \(Q_2\), the interdelivery interval
satisfies \(\interDeliv_i=H+G\), where
\(G\sim\operatorname{Geo}(Q_2p)\) on
\(\{1,2,\ldots\}\). Therefore,
\begin{align}
\mathbb{E}[\interDeliv_i]
=
H+\frac{1}{Q_2p},
\quad
\mathbb{E}[\interDeliv_i^2]
=
H^2
+\frac{2H}{Q_2p}
+\frac{2-Q_2p}{(Q_2p)^2}.
\label{eq:L1_long_pause_moments}
\end{align}
Substitution into \eqref{eq6.NAoI} with \(L=1\) gives
\begin{equation}
\widehat{\mathrm{NAoI}}_2
=
\frac{1}{2}
+
\frac{
H^2+2H/(Q_2p)+(2-Q_2p)/(Q_2p)^2
}{
2\left(H+1/(Q_2p)\right)
}.
\label{eq:L1_long_pause_AoI}
\end{equation}

\vspace{0.5ex}
\noindent\emph{Case \(L\geq2\).}
For \(L\geq2\), multiple non-tagged sources may be paused
simultaneously. As described in Sec.~\ref{secIII-A}, the
tagged source evolves over \(\mathcal{S}_2\) with transition
matrix \(\mathbf{M}_2(Q_2,p,L,H)\). The conditional pause
probability \(\pi_P^{(2)}\), conditional no-attempt probability \(Q_2\),
and stationary distribution \(\boldsymbol{\mu}^{(2)}\) are
characterized jointly by the mean-field fixed point equations
\eqref{eq:pause_probability_case2}, \eqref{eq:Q2}, and
\eqref{eq:fixed_point_case2}.

We solve the resulting fixed point system numerically. Upon
convergence, the first two interdelivery moments are computed
using
\eqref{eq:return_m1}--\eqref{eq:Ideliv2_from_TX1}, and the NAoI
follows from \eqref{eq6.NAoI}. The complete procedure is
summarized in Algorithm~\ref{alg:aoi_long_pause}.

\begin{algorithm}[tbp]
\caption{Mean-field computation of NAoI for \(H\geq L+1\)}
\label{alg:aoi_long_pause}
\begin{algorithmic}[1]

\REQUIRE
\(p\in(0,1)\), \(N\geq2\), \(L\geq2\),
\(H\geq L+1\), \(\epsilon>0\), and
\(\alpha\in(0,1]\)

\ENSURE
\(\widehat{\mathrm{NAoI}}_2\)

\STATE Initialize
\(\mu_{\mathrm{Idle}}^{[0]}\gets1\),
\(\mu_s^{[0]}\gets0\) for all \(s\neq\mathrm{Idle}\),
and \(r\gets0\)

\REPEAT

    \STATE Compute \(\pi_P^{[r]}\) from
    \eqref{eq:pause_probability_case2} using
    \(\boldsymbol{\mu}^{[r]}\)

    \STATE
    \(Q_2^{[r]}
    \gets
    [1-p(1-\pi_P^{[r]})]^{N-1}\)

    \STATE
    \(\mathbf{M}^{[r]}
    \gets
    \mathbf{M}_2(Q_2^{[r]},p,L,H)\)

    \STATE Solve
    \(\widehat{\boldsymbol{\mu}}^{[r+1]}
    =
    \widehat{\boldsymbol{\mu}}^{[r+1]}
    \mathbf{M}^{[r]}\),
    subject to
    \(\widehat{\boldsymbol{\mu}}^{[r+1]}\mathbf{1}=1\)

    \STATE
    \(\boldsymbol{\mu}^{[r+1]}
    \gets
    (1-\alpha)\boldsymbol{\mu}^{[r]}
    +
    \alpha\widehat{\boldsymbol{\mu}}^{[r+1]}\)

    \STATE
    \(d^{[r]}
    \gets
    \lVert
    \boldsymbol{\mu}^{[r+1]}
    -
    \boldsymbol{\mu}^{[r]}
    \rVert_1\)
    \STATE
    \(r\gets r+1\)

\UNTIL{\(d^{[r-1]}<\epsilon\)}

\STATE Compute \(\pi_P^\star\) from
\eqref{eq:pause_probability_case2} using
\(\boldsymbol{\mu}^{[r]}\)

\STATE
\(Q_2^\star
\gets
[1-p(1-\pi_P^\star)]^{N-1}\)

\STATE
\(\mathbf{M}^\star
\gets
\mathbf{M}_2(Q_2^\star,p,L,H)\)

\STATE Compute
\(m_1\gets\mathbb{E}[\interDeliv_i]\) and
\(m_2\gets\mathbb{E}[\interDeliv_i^2]\)
from
\eqref{eq:return_m1}--\eqref{eq:Ideliv2_from_TX1}
using \(\mathbf{M}^\star\)

\STATE
\(\widehat{\mathrm{NAoI}_2}
\gets
L+m_2/(2m_1)-1/2\)

\RETURN
\(\widehat{\mathrm{NAoI}}_2\)

\end{algorithmic}
\end{algorithm}

\subsection{Optimization of Protocol Parameters}
\label{sec:parameter_optimization}

The analytical results in \eqref{eq:NAoI_H0} and
\eqref{eq:NAoI_H_leq_L}, together with
Algorithm~\ref{alg:aoi_long_pause}, enable low-complexity
optimization of the age-threshold CSMA parameters.
The predicted NAoI depends nonlinearly on the transmission probability \(p\). In
particular, in the long-pause regime, this dependence is
implicit through the mean-field fixed point and the stationary
distribution of the tagged-source Markov chain. Since a general
convexity or unimodality guarantee is not available across all
operating regimes, we optimize \(p\) over the finite set \(\mathcal{P}
=
\{p_{\min}+m\Delta p:m=0,\ldots,M_p\}\),
where \(0<p_{\min}<1\), \(\Delta p>0\),
\(M_p\in\mathbb{Z}_{\geq0}\), and
\(p_{\max}:=p_{\min}+M_p\Delta p<1\). The resolution
\(\Delta p\) controls the tradeoff between
accuracy and computational cost.

\vspace{0.5ex}
\noindent\emph{No-pause regime.}
For \(H=0\) and \(L\geq1\), the objective is given exactly by
\eqref{eq:NAoI_H0}; denote it by
\(\mathrm{NAoI}_0(p)\). The optimal transmission probability
over \(\mathcal{P}\) can be obtained by evaluating this scalar
objective at every \(p\in\mathcal{P}\). For \(L\geq2\), a simple
closed-form optimizer over the continuous domain \(p\in(0,1)\)
is generally unavailable. For \(L=1\), however, the
continuous-domain optimizer is explicit.

\begin{corollary}
\label{cor:optimal_p_H0_L1}
For a symmetric CSMA network with \(N\geq2\), \(H=0\), and
\(L=1\), the optimal transmission probability over
\(p\in(0,1)\) and the corresponding minimum NAoI are
\begin{equation}
p_{0,\mathrm{cont}}^\star
=
\frac{1}{N},
\qquad
\mathrm{NAoI}_{0,\mathrm{cont}}^\star
=
\frac{N}{(1-1/N)^{N-1}}.
\label{eq:optimal_p_H0_L1}
\end{equation}
\end{corollary}

\begin{proof}
For \(L=1\), \eqref{eq:NAoI_H0} reduces to
\(\mathrm{NAoI}_0(p)=1/[p(1-p)^{N-1}]\).
Since \(\textstyle\frac{\mathrm{d}}{\mathrm{d}p}
\left[p(1-p)^{N-1}\right]
=
(1-p)^{N-2}(1-Np),\)
the unique maximizer of \(p(1-p)^{N-1}\) over \(p\in(0,1)\)
is \(p=1/N\), which yields \eqref{eq:optimal_p_H0_L1}.
\end{proof}

\vspace{0.5ex}
\noindent\emph{Short-pause regime.}
For each \(H\in\{1,\ldots,L\}\), the predicted NAoI
\(\widehat{\mathrm{NAoI}}_1(p,H)\) is given in closed form by
\eqref{eq:NAoI_H_leq_L}. Thus, for every fixed \(H\), the
transmission probability can be optimized by evaluating
\(\widehat{\mathrm{NAoI}}_1(p,H)\) over
\(p\in\mathcal{P}\).

\vspace{0.5ex}
\noindent\emph{Long-pause regime.}
For each \(H\in\{L+1,\ldots,H_{\max}\}\), denote the predicted
NAoI by \(\widehat{\mathrm{NAoI}}_2(p,H)\). When \(L=1\),
\(Q_2\) is obtained by solving
\eqref{eq:L1_long_pause_fixed_point}, and
\(\widehat{\mathrm{NAoI}}_2(p,H)\) is evaluated using
\eqref{eq:L1_long_pause_AoI}. When \(L\geq2\),
\(\widehat{\mathrm{NAoI}}_2(p,H)\) is returned by
Algorithm~\ref{alg:aoi_long_pause}.

Combining the three regimes, define
\begin{equation}
\widehat{\mathrm{NAoI}}(p,H)
=
\begin{cases}
\mathrm{NAoI}_0(p),
& H=0,\\
\widehat{\mathrm{NAoI}}_1(p,H),
& 1\leq H\leq L,\\
\widehat{\mathrm{NAoI}}_2(p,H),
& L+1\leq H\leq H_{\max}.
\end{cases}
\label{eq:piecewise_predicted_NAoI}
\end{equation}
For fixed \(N\geq2\), \(L\geq1\), and
\(H_{\max}\geq L+1\), the optimized protocol parameters satisfy
\begin{equation}
(p_{\rm ana}^\star,H_{\rm ana}^\star)
\in
\arg\min_{\substack{
p\in\mathcal{P}\\
H\in\{0,\ldots,H_{\max}\}}}
\widehat{\mathrm{NAoI}}(p,H).
\label{eq:joint_parameter_optimization}
\end{equation}
Algorithm~\ref{alg:parameter_optimization} summarizes the
resulting optimization procedure.

\begin{algorithm}[tbp]
\caption{Optimization of CSMA protocol parameters}
\label{alg:parameter_optimization}
\begin{algorithmic}[1]

\REQUIRE
\(N\geq2\), \(L\geq1\), \(\mathcal{P}\), \(H_{\max}\),
\(\epsilon>0\), and \(\alpha\in(0,1]\)

\ENSURE
\(p_{\rm ana}^\star,H_{\rm ana}^\star,
\widehat{\mathrm{NAoI}}^\star\)

\STATE
\(\widehat{\mathrm{NAoI}}^\star\gets\infty\)

\FOR{\(H=0,1,\ldots,H_{\max}\)}

    \FOR{\(p\in\mathcal{P}\)}

        \IF{\(H=0\)}

            \STATE Compute
            \(\widehat{\mathrm{NAoI}}(p,H)\)
            from \eqref{eq:NAoI_H0}

        \ELSIF{\(1\leq H\leq L\)}

            \STATE Compute
            \(\widehat{\mathrm{NAoI}}(p,H)\)
            from \eqref{eq:NAoI_H_leq_L}

        \ELSIF{\(L=1\)}

            \STATE Solve
            \eqref{eq:L1_long_pause_fixed_point}
            for \(Q_2\)

            \STATE Compute
            \(\widehat{\mathrm{NAoI}}(p,H)\)
            from \eqref{eq:L1_long_pause_AoI}

        \ELSE

            \STATE Set
            \(\widehat{\mathrm{NAoI}}(p,H)\)
            to the value returned by
            Algorithm~\ref{alg:aoi_long_pause}

        \ENDIF

        \IF{\(\widehat{\mathrm{NAoI}}(p,H)
        <
        \widehat{\mathrm{NAoI}}^\star\)}

            \STATE
            \((p_{\rm ana}^\star,H_{\rm ana}^\star,
            \widehat{\mathrm{NAoI}}^\star)
            \gets
            (p,H,\widehat{\mathrm{NAoI}}(p,H))\)

        \ENDIF

    \ENDFOR

\ENDFOR

\RETURN
\(p_{\rm ana}^\star,H_{\rm ana}^\star,
\widehat{\mathrm{NAoI}}^\star\)

\end{algorithmic}
\end{algorithm}

\section{Numerical Results}
\label{sec6.simulation1}

We evaluate the accuracy and computational efficiency of the
proposed AoI analysis and parameter-optimization framework.
Unless otherwise specified, \(N=10\). Each reported long
simulation is run for at least \(2\times10^7\) mini-slots and
until every source completes \(10^4\) successful deliveries.
The first \(10^6\) mini-slots are discarded as a transient
period. Reported results are averaged over ten independent runs.
The \(95\%\) confidence intervals are smaller than the plot
markers and are therefore omitted.

\subsection{Accuracy of the AoI Analysis}

\begin{figure}[!t]
\centering

\begin{subfigure}[t]{0.485\columnwidth}
    \centering
    \includegraphics[
        width=\linewidth,
        trim=2mm 2mm 2mm 2mm,
        clip
    ]{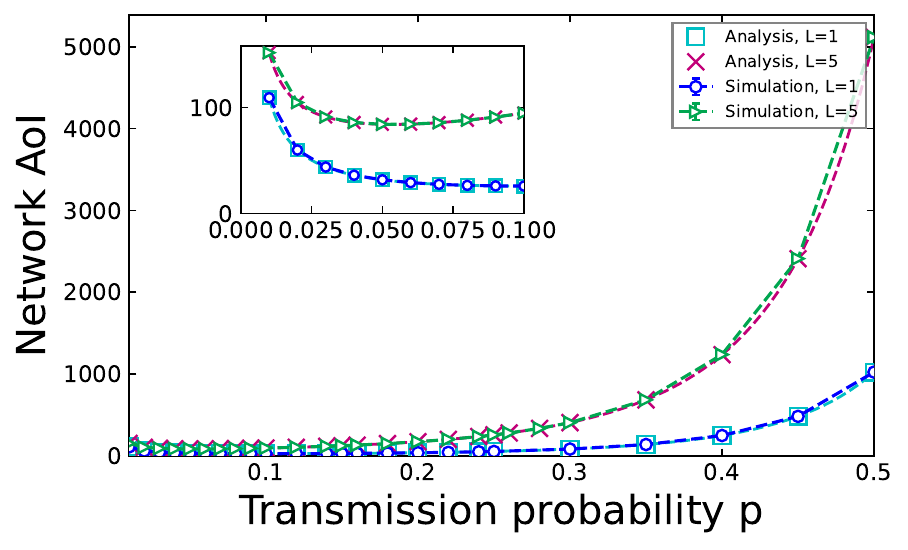}
    \caption{\(H=0\), \(L\in\{1,5\}\).}
    \label{fig:H0_validation}
\end{subfigure}
\hfill
\begin{subfigure}[t]{0.485\columnwidth}
    \centering
    \includegraphics[
        width=\linewidth,
        trim=2mm 2mm 2mm 2mm,
        clip
    ]{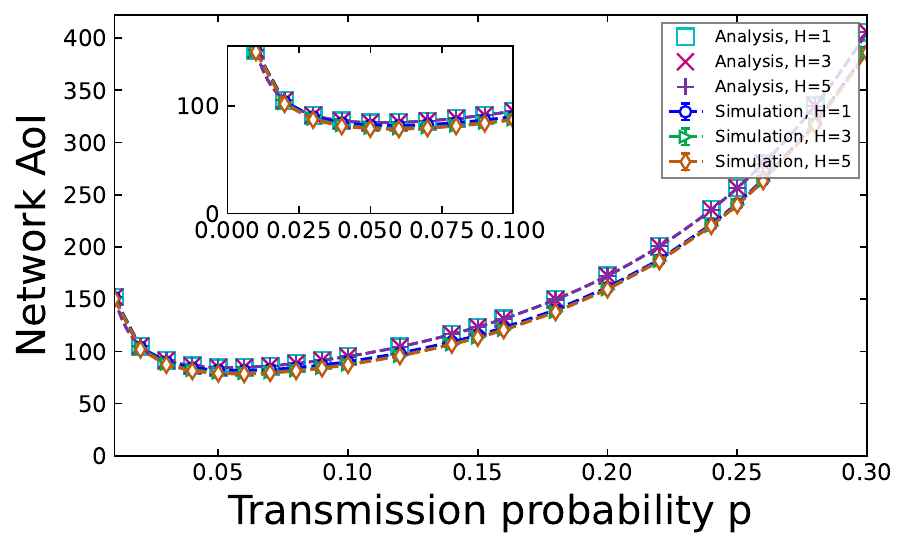}
    \caption{\(L=5\), \(H\in\{1,3,5\}\).}
    \label{fig:short_validation}
\end{subfigure}

\vspace{0.5ex}

\begin{subfigure}[t]{0.485\columnwidth}
    \centering
    \includegraphics[
        width=\linewidth,
        trim=2mm 2mm 2mm 2mm,
        clip
    ]{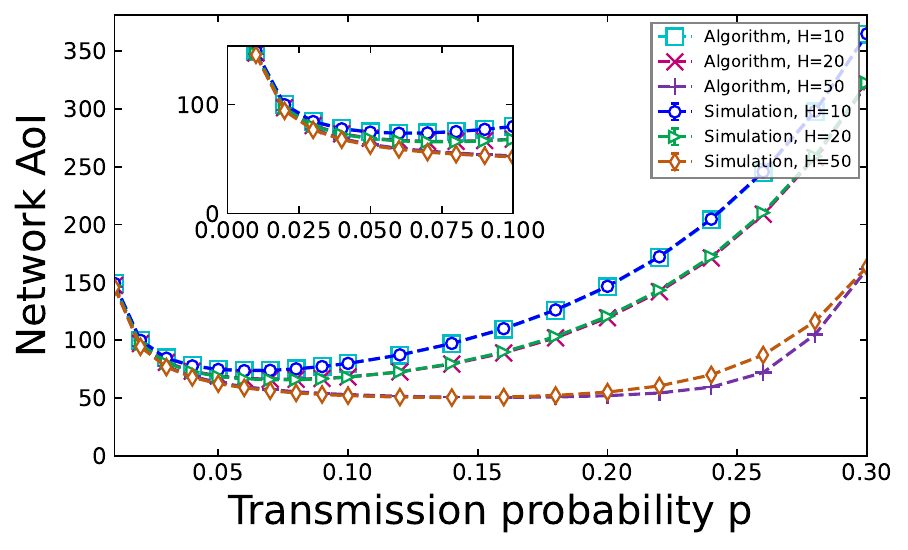}
    \caption{\(L=5\), \(H\in\{10,20,50\}\).}
    \label{fig:long_validation}
\end{subfigure}
\hfill
\begin{subfigure}[t]{0.485\columnwidth}
    \centering
    \includegraphics[
        width=\linewidth,
        trim=2mm 2mm 2mm 2mm,
        clip
    ]{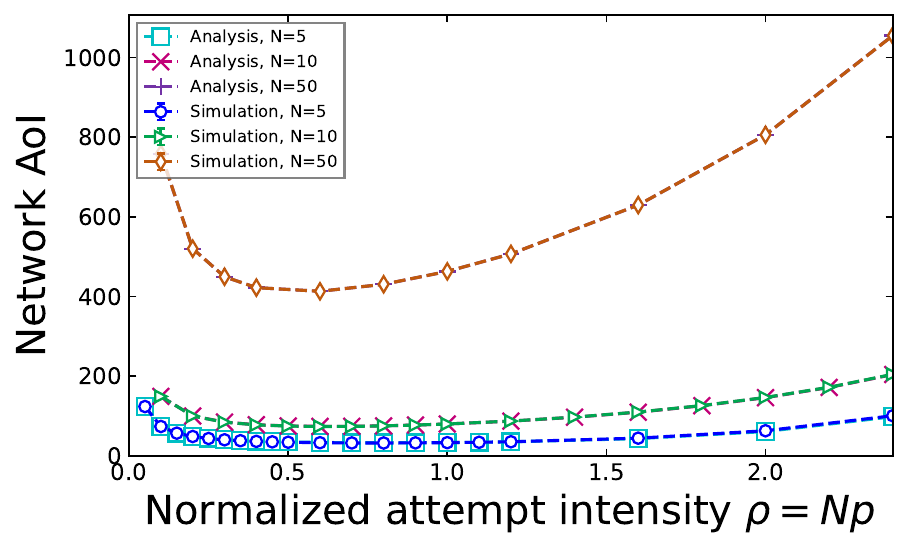}
    \caption{\(L=5\), \(H=10\).}
    \label{fig:network_size}
\end{subfigure}

\caption{Comparison of analytical and simulated NAoI.
(a) Exact no-pause analysis for \(N=10\), \(H=0\), and
\(L\in\{1,5\}\).
(b) Short-pause approximation for \(N=10\), \(L=5\), and
\(H\in\{1,3,5\}\).
(c) Long-pause approximation for \(N=10\), \(L=5\), and
\(H\in\{10,20,50\}\).
(d) Effect of network size for \(L=5\), \(H=10\), and
\(N\in\{5,10,50\}\), where \(\rho=Np\).}
\label{fig:analysis_validation}
\end{figure}

Figure~\ref{fig:analysis_validation} compares the analytical
and simulated NAoI in the three pause regimes.
Figure~\ref{fig:analysis_validation}(a) validates the exact
no-pause result in Theorem~\ref{thm:NAoI_H0}.
Figures~\ref{fig:analysis_validation}(b) and
\ref{fig:analysis_validation}(c) show that the short-pause
approximation in Theorem~\ref{thm:NAoI_short_pause} and the
long-pause fixed-point analysis in
Algorithm~\ref{alg:aoi_long_pause}, respectively, closely match
simulation. In particular, the analysis captures both the
minimum NAoI and its increase under aggressive contention.

Figure~\ref{fig:analysis_validation}(d) examines the
large-network behavior using \(N\in\{5,10,50\}\) and
\(\rho=Np\). The maximum relative error decreases from
approximately \(3.1\%\) for \(N=5\) to \(0.38\%\) for \(N=10\)
and \(0.04\%\) for \(N=50\), consistent with the large-network
mean-field approximation.

\subsection{Parameter Optimization and Computation Time}
\label{sec:optimization_results}

We compare analytical, hybrid, and exhaustive simulation-based
optimization for networks with \(N\in\{10,15,20\}\).
All three methods search over the common parameter set
\(\mathcal{P}_N\times\{0,\ldots,H_{\max}\}\).
The finite set \(\mathcal{P}_N\) covers
\(0.4\leq\rho=Np\leq2.6\), subject to \(p\leq0.25\).
For each \((N,L)\), \(H_{\max}\) is expanded whenever a
candidate optimum lies within ten slots of the current upper
boundary.

\emph{Analysis} uses
Algorithm~\ref{alg:parameter_optimization} and only the
tagged-source analytical model.
\emph{Hybrid} first obtains \(H_{\rm ana}^\star(p)\) for each
\(p\in\mathcal{P}_N\), and then runs short simulations within
\(\Delta_H=3\) of this threshold using
\(K_{\rm short}=2.5\times10^5\) mini-slots and three independent
runs. The two best \(p\)-values are retained for final long
simulations.
\emph{Simulation} performs exhaustive long-simulation search
over the same parameter set and provides the benchmark.

The analytical and hybrid gaps relative to the
simulation-search benchmark are defined as
\(\mathrm{Gap}_{\rm ana}
=(J_{\rm sim}^{\rm ana}-J_{\rm sim}^{\star})/
J_{\rm sim}^{\star}\times100\%\) and
\(\mathrm{Gap}_{\rm hyb}
=(J_{\rm sim}^{\rm hyb}-J_{\rm sim}^{\star})/
J_{\rm sim}^{\star}\times100\%\), respectively.
Among the configurations shown in
Table~\ref{tab:hybrid_comparison}, the analytical gap is at most
\(7.35\%\), while the hybrid gap is at most \(0.17\%\).
The corresponding speedups range from \(166\times\) to
\(3.26\times10^6\times\) for Analysis and from \(159\times\)
to \(934\times\) for Hybrid.
The largest analytical gap occurs at \((N,L)=(10,5)\).
This discrepancy is consistent with finite-network occupancy
fluctuations causing the mean-field model to underestimate
second-order interdelivery variability.

\begin{table}[tbp]
\caption{Optimization accuracy and computation time for
\(N\geq10\). The columns \(J_{\rm sim}^{\rm ana}\) and
\(J_{\rm sim}^{\rm hyb}\) give the long-simulation NAoI obtained
using the parameters selected by Analysis and Hybrid,
respectively. The value \(J_{\rm sim}^{\star}\) is the best
simulation result over the specified search space. Speedup is
the ratio of the exhaustive simulation-search time to the
corresponding method's optimization time.}
\label{tab:hybrid_comparison}
\centering
\setlength{\tabcolsep}{0.8pt}
\renewcommand{\arraystretch}{1.05}
\resizebox{\columnwidth}{!}{\begin{tabular}{c c c c c c c c c}
\hline
\((N,L)\)
& \((p_{\rm ana}^\star,H_{\rm ana}^\star)\)
& \((p_{\rm hyb}^\star,H_{\rm hyb}^\star)\)
& \((p_{\rm sim}^\star,H_{\rm sim}^\star)\)
& \(J_{\rm sim}^{\rm ana}\)
& \(J_{\rm sim}^{\rm hyb}\)
& \(J_{\rm sim}^{\star}\)
& \(\mathrm{Speedup}_{\rm ana}\)
& \(\mathrm{Speedup}_{\rm hyb}\)
\\
\hline
\((10,3)\) & \((0.25,41)\)  & \((0.20,41)\)  & \((0.20,41)\)  & \(34.94\) & \(33.28\) & \(33.28\) & \(8.82\times10^{3}\times\) & \(934\times\) \\
\((10,5)\) & \((0.25,62)\)  & \((0.20,62)\)  & \((0.20,64)\)  & \(52.17\) & \(48.63\) & \(48.60\) & \(506\times\)              & \(344\times\) \\
\((15,1)\) & \((0.17,27)\)  & \((0.16,28)\)  & \((0.17,30)\)  & \(23.86\) & \(23.69\) & \(23.65\) & \(1.98\times10^{6}\times\) & \(466\times\) \\
\((15,5)\) & \((0.17,98)\)  & \((0.15,99)\)  & \((0.15,99)\)  & \(74.62\) & \(71.30\) & \(71.30\) & \(166\times\)              & \(159\times\) \\
\((20,1)\) & \((0.13,37)\)  & \((0.13,38)\)  & \((0.13,39)\)  & \(31.29\) & \(31.22\) & \(31.22\) & \(3.26\times10^{6}\times\) & \(715\times\) \\
\((20,4)\) & \((0.13,110)\) & \((0.12,114)\) & \((0.11,111)\) & \(80.90\) & \(79.26\) & \(79.22\) & \(261\times\)              & \(245\times\) \\
\hline
\end{tabular}}
\end{table}

\subsection{Importance of Age-Threshold Pauses}
\label{sec:H0_vs_pause}

For \(H=0\), we compare the continuous-domain optimizer
obtained from Theorem~\ref{thm:NAoI_H0} with an independent
simulation-based optimization. As shown in
Table~\ref{tab:H0_vs_pause}, the two methods select nearly
identical transmission probabilities. The simulated NAoI at
the analytically selected parameters differs from the
simulation-based no-pause result by at most \(0.03\%\) in the
displayed configurations. Allowing \(H>0\) reduces the best
simulated NAoI over the specified joint search space by
\(40.0\%\)--\(44.2\%\) relative to the no-pause results,
demonstrating the benefit of age-threshold pauses.

\begin{table}[tbp]
\caption{Importance of age-threshold pauses and accuracy of
Theorem~\ref{thm:NAoI_H0} at \(H=0\), for \(N\geq10\).
Here, \(p_{0,\rm ana}^\star\) is obtained by continuously
minimizing the exact no-pause NAoI, whereas
\(p_{0,\rm sim}^\star\) is obtained by simulation-based
optimization with \(H=0\).
The pair \((p_{\rm sim}^\star,H_{\rm sim}^\star)\) is the best
simulation-based choice over the specified joint search space.
All \(J\) values are NAoI evaluated by long simulation, and
\(\mathrm{Improvement}
=(J_{0,\rm sim}-J_{\rm sim}^\star)/
J_{0,\rm sim}\times100\%\).}
\label{tab:H0_vs_pause}
\centering
\setlength{\tabcolsep}{1.0pt}
\renewcommand{\arraystretch}{1.05}
\resizebox{\columnwidth}{!}{\begin{tabular}{c c c c c c c c}
\hline
\((N,L)\)
& \(p_{0,\rm ana}^\star\)
& \(J_{0,\rm ana}\)
& \(p_{0,\rm sim}^\star\)
& \(J_{0,\rm sim}\)
& \((p_{\rm sim}^\star,H_{\rm sim}^\star)\)
& \(J_{\rm sim}^\star\)
& Improvement
\\
\hline
\((10,3)\) & \(0.0660\) & \(56.48\)  & \(0.0660\) & \(56.48\)  & \((0.20,41)\)   & \(33.28\) & \(41.1\%\) \\
\((10,5)\) & \(0.0536\) & \(84.10\)  & \(0.0518\) & \(84.12\)  & \((0.20,64)\)   & \(48.60\) & \(42.2\%\) \\
\((15,1)\) & \(0.0667\) & \(39.41\)  & \(0.0667\) & \(39.41\)  & \((0.17,30)\)   & \(23.65\) & \(40.0\%\) \\
\((15,5)\) & \(0.0356\) & \(126.47\) & \(0.0356\) & \(126.47\) & \((0.15,99)\)   & \(71.30\) & \(43.6\%\) \\
\((20,1)\) & \(0.0500\) & \(53.00\)  & \(0.0500\) & \(53.00\)  & \((0.13,39)\)   & \(31.22\) & \(41.1\%\) \\
\((20,4)\) & \(0.0292\) & \(141.92\) & \(0.0292\) & \(141.92\) & \((0.11,111)\)  & \(79.22\) & \(44.2\%\) \\
\hline
\end{tabular}}
\end{table}

\section{Summary}\label{sec6.summary}

In this paper, we develop a general framework for analyzing and optimizing the average AoI in Random Access networks. By combining a tagged-source mean-field model with first-return-time and renewal-reward analysis, the proposed approach provides a tractable AoI approximation without explicitly modeling the joint state of the entire network, enabling efficient optimization of the protocol parameters. The framework applies to a broad class of CSMA protocols modeled as finite-state Markov processes. As a concrete example, we study age-threshold CSMA protocols and derive closed-form or efficiently computable AoI approximations under different protocol settings.

For \(H=0\), corresponding to classical saturated CSMA, we derive an exact analytical expression for NAoI. 
For \(0<H\leq L\), we derive the first two moments of the interdelivery interval, leading to a closed-form approximation
for NAoI. For \(H\geq L+1\), we develop a mean-field fixed-point algorithm for numerically computing the approximate NAoI.

We validate the proposed analysis through extensive simulations, demonstrating close agreement between the
analytical and simulated results. 
We then use the analysis to optimize the protocol parameters to minimize AoI.

\bibliographystyle{IEEEtran}
\bibliography{references}

\end{document}